\documentclass[12pt,preprint]{aastex}

\newcommand{\be}{\begin{equation}}
\newcommand{\ee}{\end{equation}}
\newcommand{\e}[1]{\label{eq:#1}}
\newcommand{\eqref}[1]{\ref{eq:#1}}

\newcommand{\Ne}{n_{\rm e}}
\newcommand{\sigmaT}{\sigma_{\rm T}}

\newcommand{\pder}[2]{{{\partial #1} \over {\partial #2}}}
\newcommand{\Kern}{K}
\newcommand{\KKomp}{{K_{\rm K}}}
\newcommand{\Knew}{{\tilde K}} 
\newcommand{\bfKnew}{{\bf \Knew}}
\newcommand{\half}{{{1} \over {2}}}
\newcommand{\bfT}{{\bf T}}
\newcommand{\inv}{^{-1}}
\newcommand{\bfe}{{\bf e}}
\newcommand{\bfn}{{\bf n}}
\newcommand{\bfone}{{\bf 1}}
\newcommand{\bfD}{{\bf D}}
\newcommand{\bfs}{{\bf s}}

\shorttitle{New Compton Kinetic Equation}
\shortauthors{Rybicki}

\begin{document}

\title{A New Kinetic Equation for Compton Scattering}
\author{George B. Rybicki}
\affil{Harvard-Smithsonian Center for Astrophysics}
\affil{ 60 Garden Street, Cambridge, MA 02138 USA}
\email{grybicki@cfa.harvard.edu}

\begin{abstract}

A kinetic equation for Compton scattering is given that differs from
the Kompaneets equation in several significant ways.  By using an
inverse differential operator this equation allows treatment of
problems for which the radiation field varies rapidly on the scale of
the width of the Compton kernel.  This inverse operator method
describes, among other effects, the thermal Doppler broadening of
spectral lines and continuum edges, and automatically incorporates the
process of Compton heating/cooling.  It is well adapted for inclusion
into a numerical iterative solution of radiative transfer problems.
The equivalent kernel of the new method is shown to be a positive
function and with reasonable accuracy near the intitial frequency,
unlike the Kompaneets kernel, which is singular and not wholly
positive.  It is shown that iterates of the inverse operator kernel
can be easily calculated numerically, and a simple summation formula
over these iterates is derived that can be efficiently used to compute
Comptonized spectra.  It is shown that the new method can be used for
initial value and other problems with no more numerical effort than
the Kompaneets equation, and that it more correctly describes the
solution over times comparable to the mean scattering time.

\end{abstract}


\section{Introduction}\label{intro}

The most fundamental approach to treating Compton scattering of
photons from thermal electrons is to use a Boltzmann-like kinetic
equation for the photon distribution function.  The basic physics of
the Compton process is incorporated into certain scattering functions,
or kernel functions, that specify the probability of an initial photon
scattering into various frequency and angular ranges.  This Boltzmann
equation gives highly accurate results, but can also involve a heavy
computational burden, especially when the number of scatterings is
large.

The Boltzmann equation can be substantially simplified when certain
conditions are met.  Consider the case where the photons and electrons
are both non-relativistic, so that $h\nu/mc^2 \ll 1$ and $kT/mc^2 \ll
1$.  Here $h$ is Planck's constant, $\nu$ is the photon frequency, $m$
is the electron mass, $c$ is the speed of light, $k$ is Boltzmann's
constant, and $T$ is the electron temperature.  In this case the
scattering kernel is relatively narrow in frequency.  In fact the
spreading is due to the thermal Doppler width is of order $\Delta\nu
\sim \nu \alpha^{1/2}$, where,
\be 
    \alpha = {{kT} \over {mc^2}}.  \e{1.1}
\ee

When the width of the scattering kernel is small compared to the scale
over which the radiation field changes substantially, it is possible
to approximate the scattering terms in the Boltzmann equation by a
second-order differential operator acting on frequency.  Such
approximate equations are generally called Fokker-Planck equations,
but for Compton scattering the term {\em Kompaneets equation} is used,
in honor of its originator \citep{K57}.  

In its original form the Kompaneets equation applies strictly only to
homogeneous, isotropic radiation fields in which there is time
dependence but no spatial transport.  However, since the Compton
scattering process itself is not very anisotropic, the Kompaneets
scattering term is often used for non-isotropic radiation fields,
introducing some additional degree of approximation.  In this paper,
for simplicity, the transport equation is also written without the
spatial transport terms, but it should be understood that the
scattering term, within the isotropic approximation, is directly
applicable to cases involving spatial transport as well.

The Kompaneets equation has been a very useful tool for describing the
process of Compton scattering from thermal electrons in astrophysics,
when the conditions mentioned above are met.  Second-order partial
differential equations are much easier to handle numerically than full
Boltzmann equations.  Within its limitations the simple Kompaneets
equation manages to incorporate a number of important physical
properties and effects, namely,

\begin{enumerate}
  \item Conservation of photon number.
  \item Detailed balance (correct equilibrium solution).
  \item The frequency spreading due to the thermal Doppler effect.
  \item The frequency shift due to thermal Doppler effect
    (inverse Compton effect).
  \item The frequency shift due to electron recoil (Compton effect). 
  \item Stimulated scattering.
\end{enumerate}
The Kompaneets equation incorporates the first two properties exactly.  
The remaining effects are accurately
represented only to lowest order in the parameters $h\nu/mc^2$ and
$\alpha=kT/mc^2$.

For all its many positive features, however, the Kompaneets equation suffers
from one major shortcoming: it is unable to treat cases where the
radiation fields varies significantly on the scale of Compton
frequency shifts, such as can occur in the neighborhood of spectral
lines and continuum jumps.  The usual approach in such cases has been
to revert to the full Boltzmann equation, which contains the detailed
scattering kernels, or to use a Monte Carlo method.  The simplicity of
the second order differential operator is thereby lost.

An entirely different approach to the treatment of Compton scattering
was presented by \citet[ hereafter RH]{RH}, designed to apply 
to conditions typical in stellar atmospheres.  Under these
conditions the radiation fields change rapidly over the scale of the
electron thermal frequency shift $\Delta\nu$ in the neighborhood of
spectral lines and continuum edges.
The method of RH handles such rapidly varying radiation
fields in a numerically efficient way.  An essential feature of the RH
method is that it expresses the radiation field as a differential
operator acting on the emissivity, not vice versa.  This departure
from the traditional Fokker-Planck type of operator leads us to call
this an {\em inverse operator method}.

As originally formulated, the RH method incorporated only numbers 1
and 3 of the above physical effects, namely, photon conservation and
the spreading due to the thermal Doppler effect.  This was not a
serious limitation when applied to normal stellar atmospheres, where these are
the predominant effects.  However, it seemed desirable for have a
method that combined the advantages of the Kompaneets and RH
equations, which then would be applicable to a much wider class of
problems.

The purpose of this paper is to present such a new kinetic equation
for Compton scattering, which (like the Kompaneets equation)
incorporates all the physical effects 1--6 above, but which also (like
the RH method) has the ability to treat rapidly varying radiation
fields.  While the idea for this new inverse operator method was motivated
by the RH method, the derivation presented here is based on a
simple alteration of the Kompaneets equation, which is considerably
simpler than the approach used in RH.  The new method is only accurate
to first order in $h\nu/ mc^2$ and $\alpha=kT/mc^2$, and it assumes
that the scattering process can be well approximated by isotropic
emission.  However, since the Kompaneets equation itself has these
same limitations, the new method cannot be judged inferior because
of them.

From the standpoint of applications to stellar atmospheres, there are
several advantages of the new kinetic equation.  Like the Kompaneets
equation, it now incorporates the processes of Comptonization of the
radiation field and the Compton heating/cooling of the gas, which may
be of importance in certain cases involving X-ray irradiation, for
example.  Another feature is that it satisfies detailed balance, and
thus correctly describes the approach to thermal equilibrium.  At the
same time, like the RH method, it also can handle the distortion of
the line profiles due to Doppler broadening due to scattering on the
electrons.  The numerical implementation of the method involves only a
minor modification of the RH method, which essentially
maintains the latter's favorable timing requirements, namely proportional
to the first power of the number of frequency points.

For problems other than stellar atmostpheres, there are also
advantages of the new method.  For example, in studying Comptonization
of X-rays, \citet{LR79a,LR79b,LR80} wrote the solution to the transfer
problem as a sum of products of probability of scattering $k$ times
and the $k$-th iterated kernel function for Comptonization.  As we
shall see, the equivalent kernel function for the Kompaneets equation
is not wholly positive and is highly singular, being a linear
combination derivatives of delta functions up to order two.  The
``iterated'' kernels would involve even higher derivatives, and any
solution involving sums over such functions would be entirely unworkable.
On the contrary, the kernel functions of the inverse operator method
are quite normal functions, and their iterates can be found stably by
numerical means, making the method of Lightman \& Rybicki more
practical.

In \S \ref{basic} the basic properties of the Kompaneets equation will
be reviewed and the equations of the new inverse operator method will
be derived.  It is demonstrated that the inverse operator mehtod, like
the Kompaneets equation, satisfies all six of the properties listed
above.  A comparison with the RH method will be given.  In \S
\ref{kernels} we discuss properties of a rather general Boltzmann
equation and show how one may define the equivalent kernels of any
approximate version of it.  This is then done for the Kompaneets and
inverse operator methods.  Numerical results are presented for the
inverse operator kernel, and its properties, including its accuracy,
are discussed.  In \S \ref{iterated} iterates of the kernel function are
defined and discussed, and numerical examples are given.  In \S
\ref{summation} a useful formula is derived that reduces certain
summations occuring in the formalism of \citet{LR79a,LR79b,LR80} to
numerically tractable forms.  In \S \ref{iv} it is shown how initial
value problems can be treated using the inverse operator
method. Numerical results show the advantages of the inverse operator
method over the Kompaneets method for short times, of the order of the
mean scattering time.  In \S \ref{summary} a short review is given,
and some possibilities for future work are discussed.

\section{The Inverse Operator Method}\label{basic}

For isotropic homogeneous radiation, the kinetic equation for Compton
scattering can be written,
\be
   {{1} \over  {c}} \pder{n}{t} = \Ne \sigmaT(-n+e).
         \e{2.1}
\ee
Here $t$ is the time and the photon occupation number $n=n(x,t)$ is a
function of time and frequency, here given in terms of the scaled
frequency $x=h\nu/kT$.  On the right hand side $\Ne$ is the electron
density and $\sigmaT$ is the Thomson cross section, which we assume
is valid for the energies considered.  Within the
parenthesis the term $(-n)$ accounts for extinction and the term
$e=e(x,t)$ for the scattered emission.

The equation is completed by specifying a form for the scattering term $e$.  
In the well-known Kompaneets equation \citep[ \S7.6]{K57,RL}
this 
is given explicitly by,
\be
e=n+\alpha{{1} \over {x^2}}
      \pder{}{x}\left[x^4\left(\pder{n}{x}+n+n^2\right)\right].   \e{2.2}
\ee
This expression requires that the photon energy is
small compared to the electron rest energy, $h\nu/mc^2 \ll 1$,
and that the thermal electrons are nonrelativistic, $\alpha \ll
1$.  In addition, the radiation field $n(x,t)$ must vary
slowly with frequency $x$ over the typical width of the scattering
kernel, $\Delta x \sim x\alpha^{1/2}$.

The equations of the new method are found as a simple alteration of
the Kompaneets equation.  The reversion of the functional relationship
(\eqref{2.2}) between $n$ and $e$ is easily done to first order in
$\alpha$.  Since $e=n$ to lowest order, this may be used to replace
$n$ in the term multiplied by $\alpha$, giving
\be
n=e-\alpha{{1} \over {x^2}}
    \pder{}{x}\left[x^4\left(\pder{e}{x}+e+e^2\right)\right].     \e{2.4}
\ee
This implicit equation for $e$, combined with equation (\eqref{2.1}),
represents our new formulation of Compton scattering.

The Kompaneets method and the new method differ significantly in how
$e$ is determined from $n$.  In the ordinary Kompaneets equation this
is done by applying a differential operator to $n$, as in equation
(\eqref{2.2}).  However, in the new method one solves a differential
equation for $e$, using equation (\eqref{2.4}).  It might be
characterized as applying to the radiation field the inverse of a
certain operator related to the Kompaneets operator.  For this reason
we call this the {\em inverse operator method}.

The Kompaneets equation has the advantage that by substitution of
equation (\eqref{2.2}) into equation (\eqref{2.1}) one obtains a
single partial differential equation, which makes it much more amenable to
analytic treatment.  In the inverse operator method one must deal with two
separate equations, (\eqref{2.1}) and (\eqref{2.4}).  However, 
we shall show below that for many problems this pair of
equations is no more difficult to treat numerically than the
Kompaneets equation.

The inverse operator method based on equation (\eqref{2.4}) has very
different mathematical properties than the Kompaneets method.  For
example, if the radiation field $n$ varies rapidly enough, application
of the Kompaneets operator can lead to {\it negative} emission terms
$e$.  This can be seen easily from the above equations.  When the
radiation field varies significantly on the scale of the frequency
shift $\Delta x \sim x\alpha^{1/2}$, 
the second derivative term in the Kompaneets
operator become of order $\alpha^{-1} n$, so that the second term in
equation (\eqref{2.2}) is no longer of order $\alpha$, but is of order
unity.  The derivative terms, which can take either sign, are able to
dominate the first, strictly positive, term.  As we shall see below, the
inverse operator method does not lead to such negative values of $e$.

The manner of derivation of equation (\eqref{2.4}) automatically guarantees
that it has the same order of accuracy in $\alpha$ as the Kompaneets equation,
and thus satisfies all the properties 3--6 above to order $\alpha$.  
It can easily be shown that
it {\em exactly} satisfies the first two properties, namely,
photon conservation and detailed balance.

To prove photon conservation, we multiply equation (\eqref{2.4}) by
$x^2$ and integrate over all $x$.  The integrated derivative term
vanishes at the endpoints, which gives,
\be
         \int_0^{\infty} x^2 n\, dx =
         \int_0^{\infty} x^2 e\, dx.   \e{2.5}
\ee
The photon density is proportional to
\be
    {\cal  N}=  \int_0^{\infty} x^2 n \, dx.    \e{2.6}
\ee
Multiplying equation (\eqref{2.1}) by $x^2$ and integrating over all $x$,
and using equation (\eqref{2.5}), 
then yields the photon conservation result, $d{\cal N}/dt=0$.

Property 2, having the exact equilibrium solution, can be proved as follows.
In equilibrium, $\partial{n}/\partial{t}=0$, so by equation (\eqref{2.2})
we have $n(x)=e(x)$.  Substituting this into equation (\eqref{2.4})
and integrating once,
using the fact that $e(x) \rightarrow 0$ strongly as $x \rightarrow \infty$,
we find,
\be
             \pder{n}{x}+n+n^2 =0.     \e{2.7}      
\ee
The general solution to this is the general Bose-Einstein distribution,
\be
       n(x) = {{1} \over {e^{\gamma + x} -1}},  \e{2.8}
\ee
where the constant $\gamma$ is related to the chemical potential.
This is the correct photon distribution for thermal equilibrium.

Thus we have shown that the inverse operator equation (\eqref{2.4})
has the same general properties 1--6 as the Kompaneets equation.
Since the two equations differ only at a order of approximation beyond
the validity of the Kompaneets equation itself, the inverse operator
equation cannot be regarded a priori as inferior to the Kompaneets
equation.  As we shall show, the inverse operator
equation is superior in some important respects.

We conclude this section by making contact with RH.  In that work
the mean intensity $J(\nu)$ and emission integral $E(\nu)$ based on
the usual specific intensity were used.  These are related to the
quantities $n$ and $e$ by,
\be
    J(\nu)={{2h\nu^3} \over {c^2}} n(x), 
         \qquad E(\nu)={{2h\nu^3} \over {c^2}}e(x).  \e{2.9}
\ee
In terms of these variables, (\eqref{2.4}) becomes,
\be
    -\alpha \nu\pder{}{\nu} \left[ \nu \pder{E}{\nu}
  +\left( {{h\nu} \over {kT}} - 3 \right)E + {{c^2E^2} \over {2kT\nu^2}} \right]
         +E(\nu) = J(\nu).
 \e{2.10}
\ee
It is instructive to compare this to
equation (25) of RH for $N=1$, which in the present notation is
\be
    -\alpha \nu\pder{}{\nu} \left[ \nu \pder{E}{\nu} \right]
         +E(\nu) = J(\nu).
 \e{2.11}
\ee
The new inverse operator method of equation (\eqref{2.10}) is seen to be a
generalization of the RH method for $N=1$ that includes extra terms
involving first order derivatives of the radiation field.  These extra
terms will merely involve redefinitions of the coefficients of the
second-order difference equations, so, from the numerical point of
view, the new method is easily incoporated into the iterative solution
scheme given in RH.

However, now, in addition to the
treatment of Doppler broadening by scattering off thermal electrons,
all the above Compton processes 1--6, including Compton
heating/cooling, will be properly taken into account.
The new method will now be applicable to a wider range of stellar atmosphere
problems, including Comptonization of X-rays.  In cases where
the radiation should approach thermal equilibrium, this will occur
correctly.

\section{Associated Kernel Functions}\label{kernels}

Some understanding of the relationship between the Kompaneets method and
the inverse operator method can be gained by comparison	 of the equivalent
kernel functions associated with each method.  In this section we shall
define these kernel functions and give some of their general properties.
Then we shall derive explicit forms for them corresponding to the 
two methods.

The scattering process can be conveniently characterized by means of
its {\em kernel function}, also called {\em scattering function}
or {\em redistribution function}.  This function determines the probability
that an initial photon of given frequency and direction is scattered into
a final photon of some other frequency and direction.  For the simplest
case of a homogeneous, isotropic radiation field (the case treated in
this paper), the kinetic equation can be written,
\be
   {{1}\over{\Ne \sigmaT c}}\pder{n}{t} =
     \int \lbrace \left[ 1+ n(x) \right] \Kern(x,x') n(x')
  - \left[ 1+ n(x') \right] {{x'^2}\over {x^2}}\Kern(x',x) n(x) \rbrace \,dx'.
    \e{3.1}
\ee
The kernel function is denoted by $\Kern(x,x')$.  The
ratio $x'^2/x^2$, which ensures the conservation law for photons, has
its origin in certain phases space factors in the derivation (see,
e.g., Rybicki \& Lightman 1978).  Stimulated scattering is taken into
account by the factors $1+n$, which give rise to quadratic, as well as
linear, terms in the radiation field.

The integral on the right side of equation (\eqref{3.1}) must vanish
when the the radiation field is in thermal equilibrium with the
electrons, that is, when it is a Bose-Einstein distribution of the
same temperature.  The {\em principle of detailed balance} makes the
stronger statement that, in thermal equilibrium, not only must the
integral vanish, but the integrand itself must vanish frequency by
frequency.  That is,
\be
       [1+n(x)] \Kern(x,x') n(x') 
       = [1+n(x')] {{x'^2} \over {x^2}}\Kern(x',x) n(x),   \e{3.16}
\ee
for all $x$ and $x'$.  Substituting the Bose-Einstein form 
(\eqref{2.8}) into this equation leads to the condition for detailed balance,
\be
        x^2 e^{x} \Kern (x,x') = x'^2 e^{x'} \Kern (x',x).   \e{3.16.1}
\ee          

The kernel function does not depend on the radiation field $n$, so
many of its general properties can be derived by considering radiation
fields that are restricted in certain ways.  In particular, it is very
interesting to consider cases where $n \ll 1$, so that the quadratic
terms in the equation can be ignored.  The kinetic equation then
becomes,
\be
   {{1}\over {\Ne \sigmaT c}}\pder{n}{t} =
     \int \left[  \Kern(x,x') n(x')
  - {{x'^2}\over {x^2}}\Kern(x',x) n(x) \right] \,dx', \e{3.2}
\ee
which is linear in the radiation field.  Comparing this to equation
(\eqref{2.1}) we have the general normalization property,
\be
      1 = \int {{x'^2}\over{x^2}}\Kern(x',x)\,dx',  \e{3.3}
\ee
and the identification,
\be
      e(x) = \int \Kern(x,x')n(x')\,dx'.         \e{3.4}
\ee
The kinetic equation can also be written,
\be
   {{1}\over{\Ne \sigmaT c}}\pder{n}{t} 
             = -n + \int \Kern(x,x') n(x')\,dx'.  \e{3.5}
\ee

The most significant distinction between the Kompaneets method and the
new method presented here is found in the kernel functions associated
with each.  The kernel functions associated with each method are not
immediately obvious, but there is a simple trick for determining them
based on equation (\eqref{3.4}), which applies to the
case of no stimulated emission. After first changing the dummy
variable of integration to $x''$ (say), we note that by substituting
the special radiation field $n(x)=\delta(x-x')$ into the integral, we
obtain the kernel function $e(x)=\Kern(x,x')$, where $x'$ is considered
as a fixed parameter.

We shall now discuss these kernel functions, first for the
Kompaneets equation, then for the new method.  

\subsection{The Kompaneets kernel}\label{KK}

Using the trick given at the end of the previous section,
the kernel function $\KKomp(x,x')$ corresponding to the Kompaneets equation
can be found by substituting $n(x)=\delta(x-x')$ into 
equation (\eqref{2.2}) (without the stimulated term), which gives,
\be
 \KKomp(x,x') = \delta(x-x') + {{\alpha} \over {x^2}} \pder{}{x} 
 \left[ x^4 \lbrace \delta'(x-x') + \delta(x-x') \rbrace \right].   \e{3.7}
\ee
Performing the differentiation,
we can also write the Kompaneets kernel in the form,
\be
     \KKomp(x,x') = (1+4\alpha x)\delta(x-x') 
        +\alpha(x^2+4x)\delta'(x-x') + \alpha x^2 \delta''(x-x').  \e{3.8}
\ee
This shows that the Kompaneets kernel is a linear combination of
generalized functions, namely delta functions and derivatives of delta
functions up to order 2, which keep it concentrated near the point
$x=x'$.  In terms of a limiting process used to define the delta
function, such derivatives of delta functions take both
positive and negative signs, and so in a real practical sense
the Kompaneets kernel is not a completely
positive function.  This can also be verified by evaluating equation
(\eqref{3.4}) for the Kompaneets kernel with a positive, but rapidly varying,
function $n$, say a sufficiently narrow Gaussian, which will produce a
result $e$ that can take negative values. 

An important property of the kernel is its set of frequency
moments, which characterize the frequency of the scattered photon relative
to the frequency $x$ of the incident photon.  These are defined as,
\be
    \left< (x-x')^s \right> 
      = \int (x-x')^s {{x^2} \over {x'^2}} \Kern(x,x')\,dx,    \e{3.21}
\ee
where $s=0$, $1$, $2$, $\ldots$.  
The factor $x^2/x'^2$ is included to convert the kernel $\Kern$ into
the appropriate kernel for describing photon numbers, rather than
occupation number.  Substitution of the Kompaneets form for the
kernel (\eqref{3.7}) and integration by parts, one obtains the
general result,
\be
 \left< (x-x')^s \right> = \delta_{s0} +\alpha(4x'-x'^2)\delta_{s1} 
           +2\alpha x'^2 \delta_{s2}. \e{3.22}
\ee
In particular, we have,
\begin{eqnarray}
  \left< (x-x')^0 \right> &=& 1, \e{3.23.0} \\
  \left< (x-x')^1 \right> &=& \alpha (4x' -x'^2), \e{3.23.1} \\
  \left< (x-x')^2 \right> &=& 2\alpha x'^2, \e{3.23.2} \\
  \left< (x-x')^s \right> &=& 0, \qquad \hbox{\rm for $s\ge 3$}.
             \e{3.23.3}
\end{eqnarray}
The first of these is a simple restatement of the photon conservation law
for scattering.  The second and third give the results for the 
mean shift and variance of the emitted frequency relative to the
incident frequency, respectively.  These values for the moments are, of course,
well known, and were used in the derviation of the Kompaneets equation.
The above derivation simply verifies that the Kompaneets kernel
we have obtained is consistent with these values.

The final result (\eqref{3.23.3}) states that all
moments of order 3 or greater vanish.  This is a consequence of the
derivation of the Kompaneets equation, in which a Taylor series in
the emitted frequency is only carried out up to second order,
implicitly assuming that higher order moments would vanish.

\subsection{The Kernel for the Inverse Operator Method}

We have seen that the Kompaneeets equation has an equivalent kernel
that is a generalized function, containing up to second derivatives
of delta functions.  We now want to find the equivalent kernel
for the inverse operator method for treating Comptonization.  First of all
the equation for the method will be written without the stimulated
scattering terms, since, as we have seen, the kernel function can
more easily be found for this purely linear equation.  Dropping the
$e^2$ term in equation (\eqref{2.4}) gives,
\be
n=e-\alpha{{1} \over {x^2}}\pder{}{x}\left[x^4\left(\pder{e}{x}+e\right)\right].
         \e{4.1}
\ee

The kernel function determined by this equation will be denoted by
$\Knew(x,x')$.  Applying the same trick as before, we set
$n(x)=\delta(x-x')$ and $e(x)=\Knew(x,x')$ in equation (\eqref{4.1}),
which yields,
\be
\delta(x-x')=\Knew-\alpha{{1} \over {x^2}}\pder{}{x}
    \left[x^4\left(\pder{\Knew}{x}+\Knew\right)\right].
         \e{4.3}
\ee
That is, $\Knew=\Knew(x,x')$ is the Greeen's function of the differential
operator on the right hand side of equation  (\eqref{4.1}).

It is actually possible to solve this equation analytically in terms
of Bessel functions, but we omit details of this derivation.
We only quote one analytic result for very small values of $\alpha$.
Defining,
\be
\mu= \left( {{9}\over {4}} 
          + {{1}\over {\alpha}} \right)^{1/2},  \e{4.4}
\ee
the kernel is asymptotically,
\be
 \Knew(x,x') \sim {{1} \over {2 \alpha \mu}} 
     \left({{x'} \over {x^3}}\right)^{1/2}
         e^{(x'-x)/2}\left({{x_{<}} \over {x_{>}}}\right)^{\mu}. \e{4.4.1}
\ee
Here $x_{>}$ and $x_{<}$ denote the larger and smaller of $x$ and $x'$,
respectively.
Because $\mu \sim \alpha^{-1/2} \gg 1$, the last factor dominates the
behavior of the kernel, making it very sharply concentrated about
$x=x'$.  By virtue of the slowly varying factors
$(x'/x^3)^{1/2}\exp[(x'-x)/2]$, this approximate form of the kernel
still obeys the detailed balance relation (\eqref{3.16.1}).

When the kernel is sufficiently sharp, one may set $x=x'$ in the
slowly varying factors.  Taking into account the differing notations,
the kernel (\eqref{4.4}) then reduces to the form of the kernel given
in equation (17) of RH for $N=1$, showing again the relation of the
inverse operator method to RH.

It is of great practical interest here, as it was in RH, that the
scattering function $e$, or the kernel function $\Knew$, is the
solution of a second-order differential equation, (\eqref{4.1}) or
(\eqref{4.3}), respectively, since these are easily solved
numerically.  The numerical solution begins with the introduction of a
frequency grid of $N_x$ points.  Using an obvious matrix notation over
the frequency space, the Kompaneets relation (\eqref{2.2}) becomes,
\be
         \bfe = (\bfone +\alpha \bfT) \bfn,  \e{4.5}
\ee
where $\bfone$ is the unit matrix and $\bfT$ is a tridiagonal matrix 
formed from coefficients of the
discretization of the second-order differential operator.  

Similarly, the inverse operator relation (\eqref{2.4}) becomes,
\be
         \bfn = (\bfone -\alpha \bfT) \bfe,  \e{4.6}
\ee
which can be solve for $\bfe$ in terms of $\bfn$,
\be
          \bfe = (\bfone -\alpha \bfT)\inv \bfn.   \e{4.7}
\ee
Comparison of equations (\eqref{4.5}) and (\eqref{4.7}) should again 
make clear why we call this an ``inverse operator'' method.
The numerical solution of a
tridiagonal system of equations is quite rapid, with timing of order
$N_x$. This is the same
order as the multiplication of a tridiagonal matrix times a vector, so
that the numerical burden associated with the new operator for
many applications is of the
same order as that of the Kompaneets operator.  

The inverse operator kernel is similarly found by discretization of
equation (\eqref{4.3}), giving,
\be
         \bfD = (\bfone -\alpha \bfT) \bfKnew.  \e{4.8}
\ee
Here $\bfD$ is a discrete matrix version of the delta function
$\delta(x-x_0)$.  This is a diagonal matrix with diagonal elements
that depend on the quadrature scheme to evaluate integrations over
$x$ (for the crudest scheme these are equal to
the inverse of the local frequency differences).
Equation (\eqref{4.8}) can be solved column by column in order $N_x^2$
operations to give the complete inverse,
\be
          \bfKnew = (\bfone -\alpha \bfT)\inv \bfD.   \e{4.9}
\ee
If the kernel is only wanted for a few values of $x_0$, the number of operations
is correspondingly smaller.

\begin{figure}
\epsscale{0.9} 
\plotone{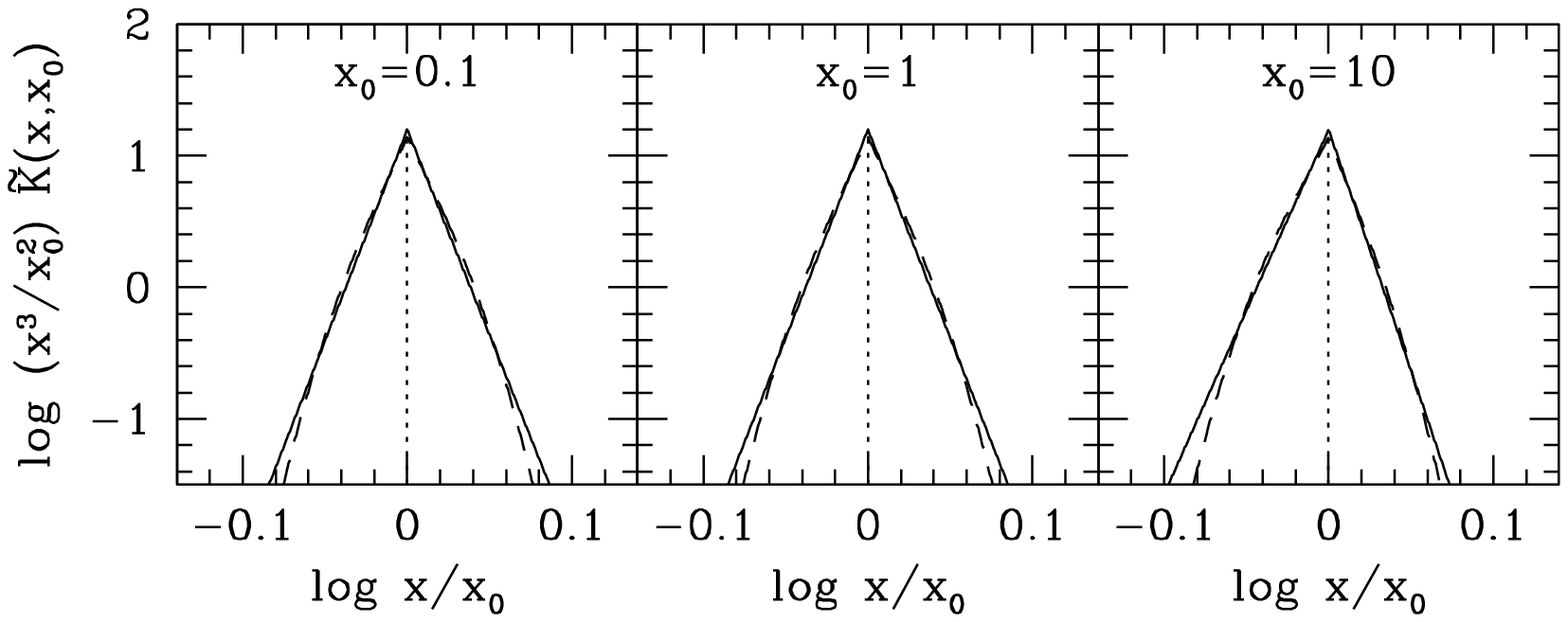}
\caption{Kernel functions for $x_0=0.1$, $1$, and $10$,
and for $\alpha=10^{-3}$.  The solid lines are
the inverse operator kernels.  The dashed curves are accurate approximations
based on expansions of \citet{Sazonov}.\label{fig:1}}
\end{figure}

In figure \ref{fig:1} results for the inverse operator kernel function
$\Knew(x,x_0)$, obtained numerically in the above way, are plotted as
the solid curves for $\alpha=10^{-3}$ and for values of $x_0=0.1$,
$1$, and $10$.  The normalizing factor $(x^3/x_0^2)$ is chosen 
to give equal areas under all the curves.
Results for the asymptotic results (\eqref{4.4}) are
also plotted as solid curves, but these are so close as to be
indistinguishable.  As expected, the kernel is skewed towards larger
(smaller) values of $x$ for $x_0\ll 1$ ($x_0 \gg 1$), expressing the
tendency for the radiation to approach thermal equilibrium.  This can
be seen most clearly for the $x_0=10$, less so for $x_0=0.1$.

In order to get some idea of the absolute accuracy of the inverse
operator results, we compare them to an expansion of the true kernel
to third order in $\alpha$ given by equation (19) of \citet{Sazonov}.
These results are plotted as dashed curves in figure
\ref{fig:1}. These show that the relative errors in the inverse
operator are less than 20\% in a central region of the kernel about
$x_0$, which extends out to values of frequency $x$ where the kernel
itself falls to a few percent of the central value.  Beyond this
central region the relative errors increase rapidly, since the true
kernel decays faster than any power law.  However, we remark that the
{\em absolute} errors outside the central region are actually quite
small.  More extensive comparisons show that the above errors are
characteristic of all the cases for which $\alpha \le 10^{-3}$ and
$x/x_0 \le 10$.

One can imagine cases where the large relative errors outside the
central region could be a problem, for instance, for accurately
describing the wings of a Compton-broadened line far away from line
center.  However, it is not common for the kernel to appear in such pure
form, and more typically the small absolute errors in
the kernel will cause lesser problems when integrations over frequency are
done; recall that the kernel does have the proper frequency moments
up to second order.

The errors in the inverse operator kernel are larger when $\alpha >
10^{-3}$, where additional relativistic corrections, omitted in the
Kompaneets equation, begin to be important.  The errors are also
larger when $x_0 \gtrsim \alpha^{-1/2}$, since the dispersion
associated with Compton recoil is not treated adequately by the
Kompaneets equation, as first discussed by \citet{Ross}.  However, the
Kompaneets equation suffers from the same problems in these regimes.

The inverse operator kernel shares some important properties with the
exact kernel: It is positive and has roughly the right shape in the
central regions, including the characteristic discontinuity of slope
for $x=x_0$.  Given that no physics beyond the Kompaneets equation has
been assumed, even the modest accuracy of the inverse operator kernels
seems most fortuitous, especially as the Kompaneets kernel itself is
singular and not wholly positive.

\section{Iterated kernels}\label{iterated}

A concept that often proves useful in practice is that of {\em
iterated kernels}.  The physical interpretation of equation (\eqref{3.4})
is that the kernel function $\Kern(x,x')$ describes how
photons of initial frequency $x'$ are redistributed to other
frequencies $x$ after one scattering.  The distribution 
after two scatterings is described by a function $\Kern_2(x,x')$, clearly
given by,
\be
       \Kern_2(x,x') = \int \Kern(x,x'')\Kern(x'',x')\,dx'', \e{6.7}
\ee
and, more generally, after $k>2$ scatterings, the distribution is described
by a function $\Kern_k(x,x')$, which is defined by the recurrence,
\be
       \Kern_k(x,x') = \int \Kern(x,x'')\Kern_{k-1}(x'',x')\,dx''.  \e{6.8}
\ee
For convenience, the definition of iterated kernels may be
extended to $k=1$ and $k=0$ by the formulas,
\be
    \Kern_1(x,x')=\Kern(x,x'), \qquad\qquad \Kern_0(x,x')=\delta(x-x'). \e{6.9}
\ee
It follows that for any nonnegative $k$ and $l$,
\be
       \Kern_{k+l}(x,x') = \int \Kern_k(x,x'')\Kern_l(x'',x')\,dx''.  \e{6.10}
\ee

These formulas may, in principle, be applied to the Kompaneets kernel
(\eqref{3.8}).  However, without working out the details, it is
obvious that the $k$-th iterated kernel is highly singular, involving
derivatives up to order $2k$ of delta functions, and is not wholly
positive.  Consequently, even for moderate values of $k$ and
relatively smooth functions, finding iterated Kompaneets kernels 
by the above recurrence relations will be impractical numerically.

For the inverse operator method, the iterated kernels, like the kernel
itself, will be ordinary positive functions, which can be easily
computed numerically.  Using the discretization introduced above,
we note that the interated kernel for the inverse operator
method can be expressed as the recurrence relation
\be
        \bfKnew_k = (1-\alpha\bfT)\inv \bfKnew_{k-1}.  \e{6.11}
\ee
The application of the inverse of a tridiagonal matrix involves of order
$N_x$ opperations, so the recurrence relation can be applied using of
order $N_x^2$ operations.  This is an order than the naive $N_x^3$ operations
if full matrices were involved.  If one only wants the $k$-th scattered
emission,
\be
         e_k(x)= \int \Knew_{k}(x,x') n(x')\, dx',  \e{6.12}
\ee
then the iterative step,
\be
        e_k(x) = \int \Knew (x,x') e_{k-1}(x')\, dx',   \e{6.13}
\ee
can be formulated, in matrix language, as,
\be
        \bfe_k = \bfKnew \bfe_{k-1} = (1-\alpha\bfT)\inv \bfe_{k-1}, \e{6.14}
\ee
which can be computed in order $N_x$ operations.

\begin{figure}
\epsscale{0.9} 
\plotone{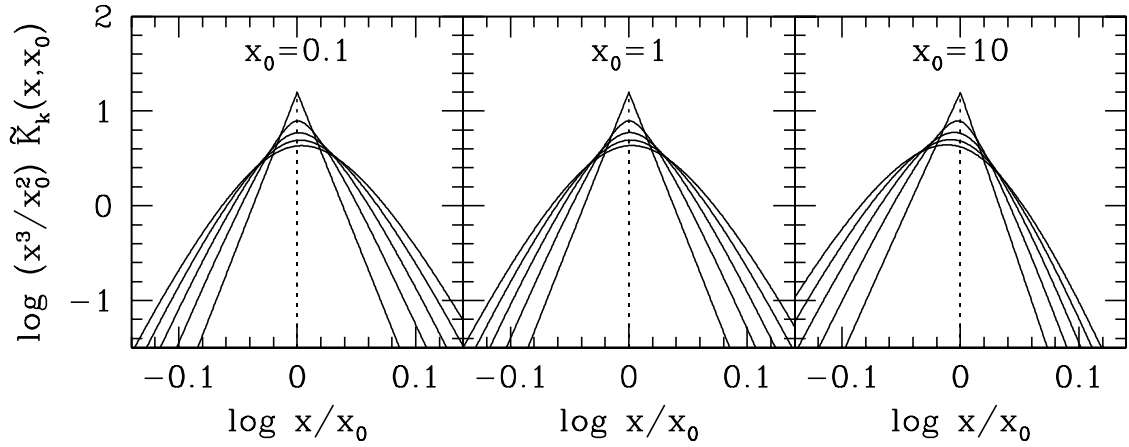}
\caption{Iterates $\Knew_k(x,x_0)$ of the inverse 
operator kernels of figure \ref{fig:1} for $k=1$, $\ldots$, 5.
The narrowest curves are the noniterated kernels ($k=1$).  The iterates
become broader as $k$ increases. \label{fig:2}}
\end{figure}

The iterates of the kernels in figure \ref{fig:1} up to $k=5$ have been
computed in this fashion and 
are shown in figure \ref{fig:2}. The discontinuity in slope of the
kernel itself disappears after one iteration, and the higher iterates
become broader and quite smooth.  The bias towards approach to thermal
equilibrium can be seen in the tendency for the higher iterates to
move toward the central Wien frequency of $x \sim 3$.

Repeated application of equations (\eqref{6.11}) and (\eqref{6.14}) lead to
the explicit forms,
\be
   \bfKnew_k = (\bfone-\alpha \bfT)^{-k} \bfD, \e{6.15}
\ee
and
\be
    \bfe_k = (\bfone-\alpha \bfT)^{-k} \bfe_0, \e{6.16}
\ee
in terms of powers of the matrix $(\bfone-\alpha\bfT)^{-1}$.

\section{A Summation Formula}\label{summation}

Equation (\eqref{6.16}) is the basis of a formula we have found very useful
for the computation of Comptonized spectra using the formalism of
\citet{LR79a,LR79b,LR80}.  Putting that formalism into the present notation,
a spectrum $s(x)$ is expressed as the sum,
\be
             s(x) = \sum_{k=0}^\infty p_k e_k(x),   \e{6.17a}
\ee
where $p_k$ is the probability of a photon in the spectrum having scattered
$k$ times before escaping.  Again, in matrix form,
\be
             \bfs =\sum_{k=0}^\infty p_k \bfe_k.   \e{6.17b}
\ee

It turns out that $p_k$ can often be accurately expressed as the
linear combinations of exponential terms of the form $A z^{-k}$, where
$A$ and $z>1$ are constants.  We shall derive a formula for the
spectrum associated with just one term, since the large-$k$ behavior of
$p_k$ is often dominated by one term; examples of this can be found in
\citet{Nishimura}.  In any case, by linear superposition, we can use
the formula to find the spectrum due to a sum of such terms.

Thus we consider the exponential expression for $p_k$,
\be
        p_k = A z^{-k} = (1-z^{-1}) z^{-k},   \e{6.18}
\ee
where for convenience we take $A=(1-z^{-1})$ so that the $p_k$ are normalized,
\be
           \sum_{k=0}^\infty p_k=1.           \e{6.19}
\ee

The summation (\eqref{6.17b}) for this
choice of $p_k$ and with equation (\eqref{6.16}) is,
\be
  \bfs 
=A \sum_{k=0}^\infty \left[ z^{-1}(\bfone-\alpha\bfT)^{-1} \right]^k \bfe_k
= A\left[ \bfone -z^{-1}(\bfone-\alpha\bfT)^{-1} \right]^{-1} \bfe_k,  \e{6.20}
\ee
summing the geometric series.  A slight rearrangement
yields the desired formula,
\be
  \bfs = A\bfe_0 + Az^{-1}(\bfone -\alpha\bfT-z^{-1})^{-1}\bfe_0. \e{6.21}
\ee
The second term may be found as the solution of a tridiagonal system
in order $N_x$ operations, so the computation is very fast.

\begin{figure}
\epsscale{0.8} 
\plotone{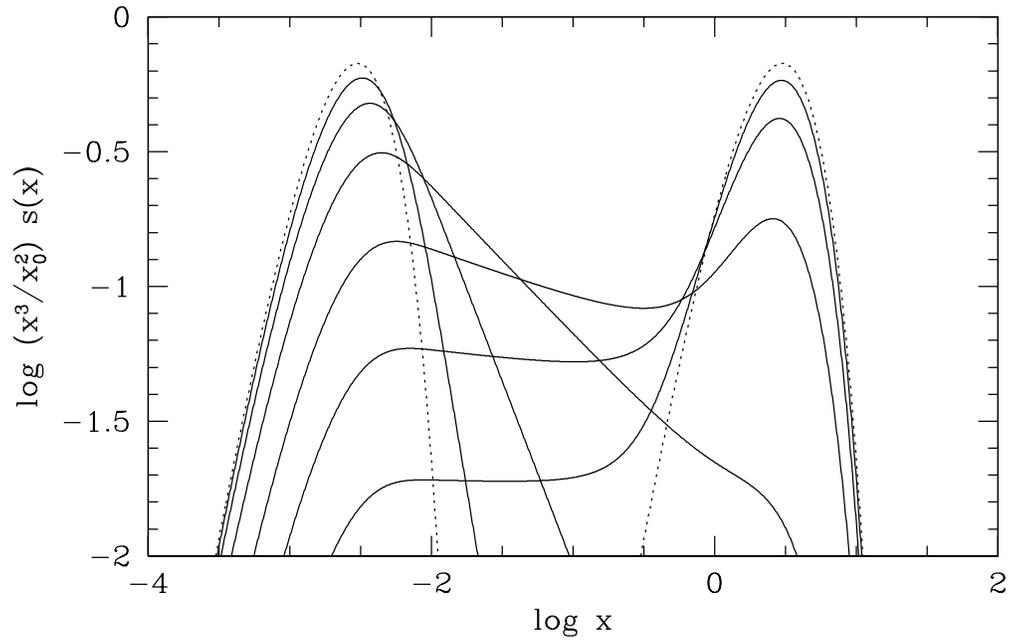}
\caption{Comptonized spectra from a soft initial Wien source
($T_0=10^4$ K) scattering in a hot Comptonizing medium ($T=10^7$ K)
with simple model for the escape probability (see text).
The solid curves, progressing from left to right, are for values of $y_\star=$
0.03, 0.1, 0.3, 1, 3, and 10.  The Wien curves for $10^4$ K and $10^7$ K
are shown as dotted on the left and right, respectively. \label{fig:3}}
\end{figure}

Examples of spectra computed using our summation formula are given in
figure \ref{fig:3} for a source $e_0(x)$ of soft photons, 
emitted in a Wien law of temperature $T_0=10^4$ K,  which are Comptonized
by scattering in a hot medium of temperature $T=10^7$ K, making
$\alpha=1.68 \times 10^{-3}$.  The choice of normalizing factors here
for $s(x)$ makes the plot analogous to a $\log \nu F_\nu$ vs.\ $\log \nu$ 
plot, where $F_\nu$ is the photon flux.  The values
of $z$ are conveniently characterized by the parameter,
\be
         y_\star = \alpha/\ln z,     \e{6.22}
\ee
which is a kind of overall ``$y$-parameter'' for the scattering process.  To see
this, we examine the limit where $z$ is close to unity, so that
$\ln z \approx z-1$.  But the mean number of scatterings is
 $\langle N \rangle =\sum_k kp_k = (z-1)^{-1}$ from 
equation (\eqref{6.18}), so that in the limit we have,
\be
         y_\star \sim \alpha \langle N \rangle,  \e{6.23}
\ee
which is a typical definition for a $y$-parameter; see, e.g.,  
\citet[ Eq.\ 7.41a]{RL}.  In figure \ref{fig:3}, as the values of
$y_\star$ range from 0.03 to 10, the spectrum changes from being nearly
a Wien law at the initial $10^4$ K temperature to being nearly a Wien law
at the temperature $10^7$ K of the Comptonizing medium.  For values
of frequency between these two Wien laws, one sees an approximate power law,
which is originially steep, but which eventually becomes flat.

\section{Initial value problems}\label{iv}

In an initial value problem the radiation field is specified at an
initial time, say $t=0$, and the problem is to find its subsequent
behavior as a function of frequency and time, that is, given $n(x,0)$,
find $n(x,t)$.  In this section we shall describe the results of
numerical solutions of initial value problems for the Kompaneets and
the inverse operator methods for the spatially homogeneous case.  This
will help clarify some of the distinctions between the two methods.

For simplicity the stimulated scattering process will be neglected, so
that the equations are linear.  The prototypical initial value problem
is to start with an initial radiation field concentrated at a single
frequency $x_0$, so that $n(x,0)=\delta(x-x_0)$.  The
discretized vector version of this initial function will be denoted
$\bfn_0$.  Any initial value
problem can then be solved by linear superposition of such solutions.

It is convenient to replace $t$
by the {\em Compton $y$-parameter},
\be
       y=\alpha\tau = \left({{kT} \over {mc^2} }\right) \Ne\sigmaT c t,  \e{5.1}
\ee
which measures time in units of the time for significant
Comptonization to take place.
Combining equations (\eqref{2.1}) and (\eqref{2.2}) and writing the
result in matrix form, we then have,
\be
        \pder{\bfn}{y} = \bfT\bfn.  \e{5.2}
\ee
This equation can be solved by the well-known Crank-Nicholson method.
Introducing a discretization in $y$ (time), indicated by a subscript $j$,
we write,
\be
        {{\bfn_{j+1}-\bfn_{j}} \over {\Delta y}} = \half \bfT \bfn_{j}
                 +\half \bfT \bfn_{j+1} .   \e{5.3}
\ee
That is, we use a differencing scheme that is semi-implicit in time.
This can be rearranged into the form,
\be
        (\bfone-\Delta_0 \bfT) \bfn_{j+1} 
     = (\bfone+\Delta_0 \bfT) \bfn_{j},  \e{5.4}
\ee
where, 
\be
       \Delta_0=\Delta y /2.   \e{5.5}
\ee

For the initial value problem we are given the radiation field
$\bfn_0$ at $y=0$, and we construct the $\bfn$'s at succeeding times
by using (\eqref{5.5}) as a recurrence relation.  Given the solution
$\bfn_j$ we evaluate the right hand side (this involves multiplying by
a tridiagonal matrix) then solving the resulting equation for
$\bfn_{j+1}$ (this involves solving a linear system with tridiagonal
matrix).  Both operations involving tridiagonal matrices requires only
$N_x$ operations, so they are very rapid.

Now let us turn to the initial value problem using the inverse operator
method.  Using the same discretization, equations (\eqref{2.1})
and (\eqref{2.4}) become,
\be
      \pder{\bfn}{y} = \alpha\inv\left[-\bfn +(1 - \alpha \bfT)\inv \bfn \right]
                        = (\bfone - \alpha \bfT)\inv \bfT \bfn.  \e{5.6}
\ee
This can be differenced using the Crank-Nicholson method, as before,
so that,
\be
        {{\bfn_{j+1}-\bfn_{j}} \over {\Delta y}} 
                  = \half (\bfone - \alpha \bfT)\inv\bfT \bfn_{j}
                 +\half (\bfone - \alpha \bfT)\inv\bfT \bfn_{j+1} .   \e{5.7}
\ee
After multiplication by $(\bfone - \alpha \bfT)$, this can be rearranged 
into the form,
\be
        (\bfone-\Delta_{+} \bfT) \bfn_{j+1} 
         = (\bfone+\Delta_{-} \bfT) \bfn_{j},  \e{5.8}
\ee
where,
\be
       \Delta_{+}=\Delta y /2+\alpha, \qquad\qquad
       \Delta_{-}=\Delta y /2-\alpha.   \e{5.9}
\ee
Equation (\eqref{5.8}) is a trivial modification of (\eqref{5.4}), the
only change being the replacement of the two appearances of $\Delta_0$
with $\Delta_{+}$ and $\Delta_{-}$ on the opposite sides of the
equation.  Thus the numerical method developed for the Kompaneets
equation can be trivially adapted to solve the initial value for the
inverse operator method, and it will be equally fast.

One distinction between the two methods is now apparent.
For the Kompaneets equation, the time variable $t$ and 
$\alpha$ can be combined into a single variable, the $y$-parameter.
This implies that all initial value problems starting from the same
initial radiation field will have exactly the same evolution for
different values of $\alpha$, except for a simple rescaling of the
time variable.  For the inverse operator method, even using the
$y$-parameter does not completely eliminate the dependence on
$\alpha$, as can be seen from equation (\eqref{5.6}), or from the
fact that $\alpha$ appears in the definitions of $\Delta_{+}$ and
$\Delta_{-}$.  This implies that the inverse operator method 
describes the solution
not only on the time for significant Comptonization, but also
on the typically much shorter mean free time for scattering.

\begin{figure}
\epsscale{0.8} 
\plotone{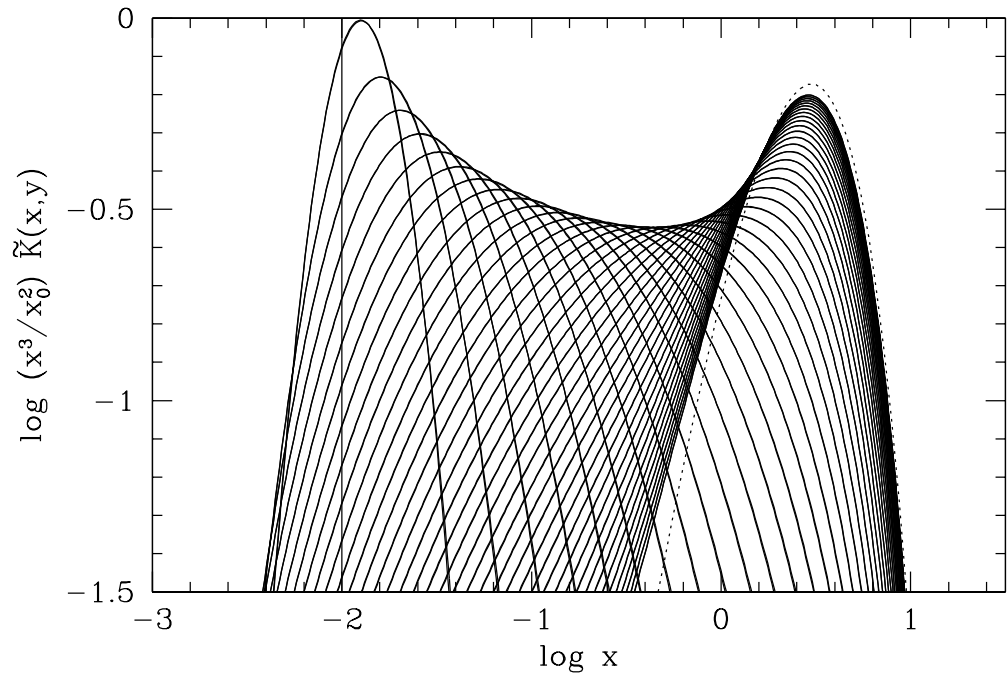}
\caption{Solutions (solid) for the initial value problem with an
initial delta-function at $x_0=10^{-2}$ (vertical line) and for 
$y=0$, 4(0.1). For large values of $y$ the solutions approach the
Wien law (dotted).  The Kompaneets and 
inverse operator solutions are virtually indistinguishable in this plot.
\label{fig:4}}
\end{figure}

An example of the numerical solution of initial value problems is
shown in Figure \ref{fig:4}.
Here $\alpha=10^{-3}$, and the initial radiation field is a
delta-function at the frequency $x_0=10^{-2}$.  The values of the
$y$-parameter range from $0$ to $4$ in steps of $0.1$.  For the larger
values of $y$, the solution is seen to be converging to the Wien law,
$(1/2)x^3\exp(-x)$, shown as the dotted curve.  
Both the Kompaneets and the inverse
operator results are plotted in figure \ref{fig:4}, but for this set
of $y$ values the two solutions cannot be distinguished to plotting
accuracy.

The discussion above suggests that the distinctions between the two
methods will become manifest for times comparable to the mean free
scattering time, $t \lesssim (\Ne\sigmaT c)^{-1}$, that is, when $\tau
\lesssim 1$ or $y \lesssim \alpha$. 

\begin{figure}
\epsscale{0.7} 
\plotone{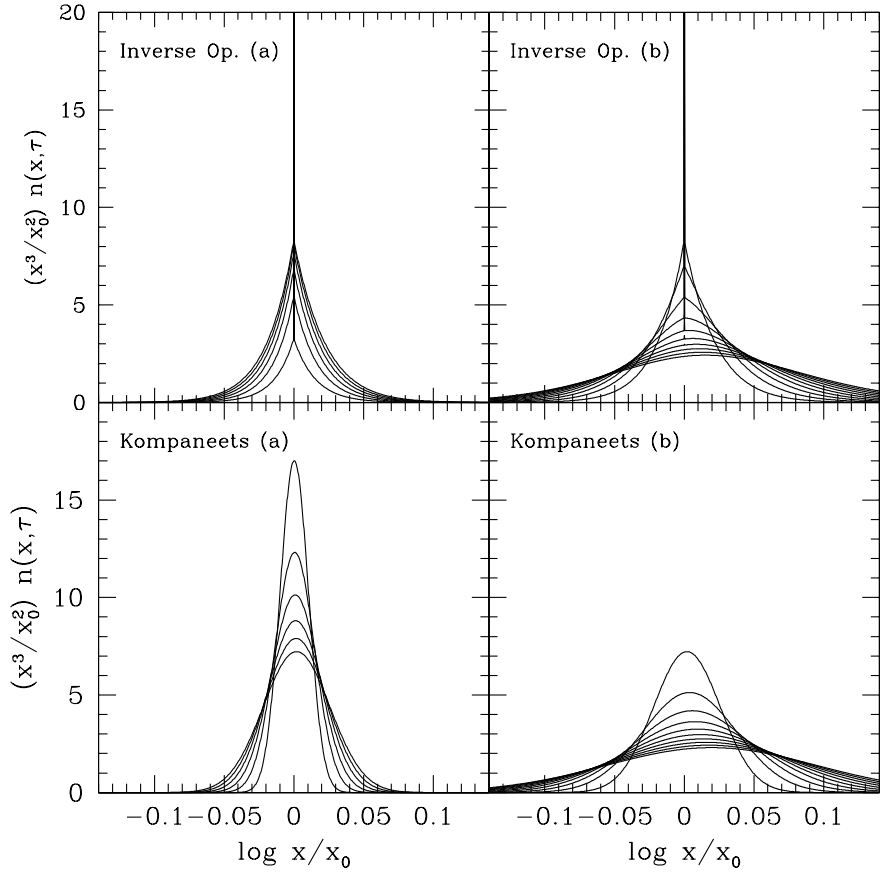}
\caption{Short time behavior of the initial value problem with 
$\alpha=10^{-3}$ and an
initial delta-function at $x_0=10^{-2}$ for the inverse operator method (upper
panels) and the Kompaneets equation (lower panels).  Values of $\tau$
are (a): 0.25,1.5(0.25) for the left panels and (b): 1.5,15(1.5) for the right
panels.  Note the linear vertical scale here. \label{fig:5}}
\end{figure}

Figure \ref{fig:5} demonstrates that the behaviors of the two methods
at relatively short times are indeed very different.  The upper set of
curves show the inverse operator results and the lower set the
Kompaneets results.  To avoid confusing overlapping curves, the times
have been divided into two sets, $\tau=0.25$ to $1.5$ in steps of
$0.25$ for the left panels (a), and $\tau=1.5$ to $15$ in steps of
$1.5$ for the right panels (b).  The initial radiation field $n(x,0)=
\delta(x-x_0)$ with $x_0=10^{-2}$ has not been plotted here, but would 
have appeared as a
vertical line in all plots.  

For the inverse operator method, one component of the
solution is clearly a remnant of the localized
initial delta function.  Immediately
outside this, there is an extended component that increases with time until
$\tau \sim 1.5$, then falls.  The extended component originally has
a noticeable discontinuity of slope at $x=x_0$, but eventually becomes
smoother as it broadens.  

This behavior of the inverse operator results is easily understood analytically.
One shows by direct substitution that the solution to the
time dependent equation (\eqref{3.5}) with $n(x,0)=\delta(x-x_0)$ can
be written in terms of iterated kernels,
\be
    n(x,\tau) = \sum_{k=0}^{\infty} {{\tau^k e^{-\tau}} \over {k!}} K_k(x,x_0).
          \e{5.10}
\ee
For moderate times
$\tau \lesssim 1$, we expect only a few terms of this series to
contribute, so that
\be
    n(x,\tau) = e^{-\tau}\delta(x-x_0) + \tau e^{-\tau} K(x,x_0) + \ldots.
   \e{5.11}
\ee

Thus, for the inverse operator method, we expect an early time behavior of a
decaying delta function plus a rising contribution of the noniterated kernel,
peaking at $\tau = 1$, then decaying away.  The presence of the
delta function expresses the fact that after a time $\tau$, some
fraction (in fact, $e^{-\tau}$) of the photons have not yet scattered;
in principle, this delta function never goes away, but in the
numerical solution it eventually becomes negligible.  The presence of
the noniterated kernel can be recognized by the discontinuity in slope
at $x_0$, but at later times this feature decays and disappears as further terms
involving iterated kernels begin to dominate.  This is precisely what
is seen in the numerical solutions for the inverse operator method
in the upper panels of figure \ref{fig:5}.  

Equation (\eqref{5.10}) is not a useful representation of the solution
of the Kompaneets equation, due to the singular nature of the iterated
kernels, so the preceding discussion is not applicable to it.  The
numerical solution of the Kompaneets shown in the lower panels of
figure \ref{fig:5} never exhibits a trace of a delta function for
$\tau > 0$, but rather shows quasi-Gaussian forms, which continually
broaden and shift with time.  For the largest time ($\tau=15$) the
results are virtually indistinguishable from those of the inverse
operator method.

The coefficient of the iterated kernel in the representation
(\eqref{5.10}), that is, $\tau^k e^{-\tau}/k!$, has the same form as
the Poisson probability distribution, and for large $\tau$ (or large
$k$) is sharply peaked about $k=\tau$.  This suggests the large-$k$
asymptotic approximation to the iterated kernels,
\be
        K_k(x,x_0) \sim n(x,\tau=k),  \e{5.12}
\ee
where $n(x,\tau)$ is the initial value problem with $n(x,0)=\delta(x-x_0)$.
Indeed, this is essentially the basis of the method of \citet{ST}, who
used it to determine Compton spectra.  

For small $\tau$ the mixing of the different values of $k$ in equation
(\eqref{5.10}) is significant, and the approximation (\eqref{5.12}) is
not so good.  Nonetheless, \citet{Sunyaev}
suggested using equation (\eqref{5.12}) 
for $k=1$ to define the approximate kernel,
\be
        K_{\rm SS}(x,x_0)= n(x,\tau=1),  \e{5.13}
\ee
which he called
the ``Kompaneets kernel.''  This differs from our
``Kompaneets kernel'' $\KKomp$, which is the true kernel of the
Kompaneets equation, as given in equation (\eqref{3.8}).  The kernel
$K_{\rm SS}(x,x_0)$ does have the desired property that the $k$-th
iterate of it is exactly equal to $n(x,\tau=k)$, so it does give
asymptotically accurate iterates through equation (\eqref{5.12}).
However, we find that $K_{\rm SS}$ is not as accurate as the inverse
operator kernel in the central regions, and it does not have the
characteristic discontinuity of slope at $x=x_0$.  In addition,
$K_{\rm SS}$ is actually harder to compute than the inverse operator
kernel, since it is defined by an integration of the time-dependent
Kompaneets equation over a finite interval of time.

\section{Summary and Discussion}\label{summary}

We have demonstrated a simple modification of the Kompaneets method
that maintains most of its desirable properties, but which has much
better behavior at very short times, of order of the mean free
scattering time.  This inverse operator method has an equivalent
kernel function that is positive and nonsingular, unlike that of the
Kompaneets equation.  The kernel function is roughly of the right
shape in the central regions with an accuracy of about 20\%.  We have
shown that for many applications the inverse operator method requires
no more numerical effort than the Kompaneets equation.

We have shown how the iterated kernels of the inverse operator method
can be efficiently computed numerically.  A summation formula
involving iterated kernels is given that reduces the effort in computing
Comptonized spectra using the method of \citet{LR79a,LR79b,LR80}.  
We have shown that initial value problems can be solved using the
inverse operator method with no more effort than the Kompaneets equation,
but with noticeable improvement in the solution at times comparable
to the mean scattering time.

In this paper there has been no attempt to advance beyond the physics
of the Kompaneets equation.  The main goal has been simply to point
out the special advantages that occur when one reverses the role of
the second order differential operator that connects the radiation
field with the emission coefficient.  However, in future work it would
be desirable to overcome some of the limitations of the inverse
operator method, as others have done for the Kompaneets equation.
There does not seem to be any reason to expect any difficulties in
extending the present theory to anisotropic scattering or
polarization.

It would be desirable to find ways to increase the accuracy of the
inverse operator method.  Obvious improvements would come with using
higher order expansions of the coefficients in the parameters $\alpha$
and $h\nu/mc^2$, but this will not solve all of the accuracy issues.

More improvement might be gained by expressing the emission
coefficient as a sum of $N>1$ terms, as in RH.  Alternatively, and
more in keeping with the present approach, one could include higher
order moments and derivatives in a Kramers-Moyal expansion \citep[see, e.g.,
][]{Riskin} of the Boltzmann equation, and then use this to form new,
higher order, inverse operators.  

The present work has concentrated on one particular Fokker-Planck
equation, the Kompaneets equation.  However, the idea of using inverse
operators clearly could be generalized and applied to other
Fokker-Planck equations.  Since each physical situation has its own
special properties, it is not possible to predict whether an inverse
operator approach would provide significant advantages in any particular
case, but it might be worth investigating.  An intriguing
possibility is that higher order Kramers-Moyal expansions might
have better properties when modified by the use of inverse
operators, which perhaps might avoid some of the difficulties known to
exist for the traditional expansions beyond the Fokker-Planck approximation
\citep{Pawula}.

\acknowledgements

The author wishes to thank A.\ Loeb and R.\ Narayan for helpful
discussions and comments.  The paper was also improved through the
valuable suggestions of S.Y.\ Sazonov, D.G.\ Hummer, and especially
the referee, I.\ Hubeny.

\clearpage

\clearpage

\clearpage

\clearpage

\clearpage

\clearpage

\begin{thebibliography}{}

\bibitem[Kompaneets(1957)]{K57}
Kompaneets, A.S., 1957, Sov. Phys. JETP, 4, 730

\bibitem[Lightman \& Rybicki(1979a)]{LR79a}
Lightman A.P., \& Rybicki G.B.  1979a, \apjl, 229, L15

\bibitem[Lightman \& Rybicki(1979b)]{LR79b}
Lightman A.P., \& Rybicki G.B.  1979b, \apj, 232, 882

\bibitem[Lightman \& Rybicki(1980)]{LR80}
Lightman A.P., \& Rybicki G.B.  1980, \apj, 236. 928

\bibitem[Nishimura et al.(1986)]{Nishimura}
Nishimura, J., Mitsuda, K., \& Itoh, M. 1986, \pasj, 38, 819

\bibitem[Pawula(1967)]{Pawula}
Pawula, R.F. 1967, Phys. Rev., 162, 186

\bibitem[Riskin(1984)]{Riskin}
Riskin, H. 1984, The Fokker-Planck equation: methods of solution and 
  applications, Springer-Verlag, New York

\bibitem[Ross et al.(1978)]{Ross}
Ross, R.R., Weaver, R., \& McCray, R. 1978, \apj, 219, 292

\bibitem[Rybicki \& Hummer(1994)]{RH}
Rybicki, G.B., \& Hummer D.G., 1994, \aap, 290, 553 (RH)

\bibitem[Rybicki \& Lightman(1979)]{RL}
Rybicki G.B., \& Lightman A.P., 1979, Radiative Processes in
 Astrophysics, John Wiley \& Sons, New York

\bibitem[Sazonov \& Sunyaev(2000)]{Sazonov}
Sazonov, S.Y., \& Sunyaev, R.A. 2000, \apj, 543, 28

\bibitem[Sunyaev(1980)]{Sunyaev}
Sunyaev, R.A. 1980, Sov. Astron. Lett., 6(4), 213

\bibitem[Sunyaev \& Titarchuk(1980)]{ST}
Sunyaev, R.A., \& Titarchuk, L.G. 1980, \aap, 86, 121

\end{thebibliography}
\end{document}